\documentclass[lettersize,journal]{IEEEtran}
\usepackage{amsmath,amsfonts}
\usepackage{algorithmic}
\usepackage{algorithm}
\usepackage{array}
\usepackage[caption=false,font=normalsize,labelfont=sf,textfont=sf]{subfig}
\usepackage{textcomp}
\usepackage{stfloats}
\usepackage{url}
\usepackage{verbatim}
\usepackage{graphicx}
\usepackage{cite}
\usepackage{epstopdf}
\usepackage{multirow}
\usepackage{multicol}
\usepackage{makecell}
\usepackage{tabularx}
\usepackage{amsmath}
\usepackage{amssymb}
\usepackage[table,xcdraw]{xcolor}
\usepackage{booktabs}
\usepackage{multirow}
\usepackage{graphicx}
\usepackage{xcolor}

\begin{document}
\title{zSort: Stable Distribution Sort using Z-Score Partitioning}
\author{Hriday Jain, Ketan Sabale, Aditya Shastri, Hiren Kumar Thakkar and Ashutosh Londhe
	\thanks{Hriday Jain, Ketan Sabale, Aditya Shastri and Hiren Kumar Thakkar are with the Department
		of Computer Science and Engineering, Pandit Deendayal Energy University, Gujarat, India. 
		E-mail: (hridayjain118@gmail.com, ketansabale1@gmail.com, Aditya.Shastri@sot.pdpu.ac.in, Hiren.Thakkar@sot.pdpu.ac.in)}
	\thanks{Ashutosh Londhe is with the School of Mechanical and Aerospace Engineering
		Queen's University Belfast, United Kingdom  
		E-mail: (a.londhe@qub.ac.uk)}}
\maketitle
\begin{abstract}
Sorting is a foundational primitive in modern data processing, influencing the execution speed of high-performance data pipelines. However, the algorithmic landscape is currently bifurcated by a pervasive ``Stability Tax": practitioners must sacrifice either order preservation for high throughput  or execution speed for stability. To address these limitations, this paper introduces, zSort, an adaptive z-score based distribution sorting algorithm that guarantees stability while avoiding pass complexity that scales with key-width. The performance of the proposed technique is evaluated using Microarchitectural analysis and experimental results. Microarchitectural analysis shows that zSort achieves a lower bad-speculation overhead (19.7\%) than both stable baselines and several high-performance unstable algorithms and sustains a competitive IPC of 1.44. Empirical evaluation across diverse input distributions and datasets of up to $10^7$ elements (64 bit) demonstrates that zSort consistently outperforms widely used comparison based stable sorting algorithms, achieving up to 3$x$-4.5$x$ speedups, and a relatively better performance compared to LSD Radix, with larger gains on duplicate heavy and partially ordered inputs. Despite providing stability, zSort achieves comparable throughput as compared to high-performance unstable algorithms such as Skasort. It also maintains this performance on adaptive workloads where methods like Pdqsort typically excel and doesn't exhibit any extreme worst case. These results indicate that zSort substantially narrows the traditional performance gap between stable and unstable sorting and provides an efficient, stable sorting alternative. 
\end{abstract}


\section{Introduction}
\IEEEPARstart{I}{n} the landscape of modern computing, spanning from high-performance database engines and operating system schedulers to large-scale data analytics, the efficient organization of data remains to be a fundamental operation. Often perceived as a basic utility, sorting is a fundamental subroutine that significantly influences the throughput of real-world applications \cite{1.1}. Moreover, the continuous evolution of sorting algorithms serves as a driver for advancements in the broader field of Algorithm Engineering.\par 
Despite the theoretical maturity of the field, the practical cost of sorting in real-world infrastructure remains non-trivial and often underestimated. The author in \cite{1} establishes that sorting algorithm is often the dominant cost factor in modern database systems. It is used in variety of tasks including query processing, object
assembly and record access, index creation and maintenance, and consistency checks. The industry’s reliance on sorting as a proxy for overall system performance is evidenced by the GraySort Benchmark \cite{2}, a competition in which large-scale systems optimize their hardware stacks to maximize sorting throughput, recognizing it as a stress test for memory bandwidth and I/O efficiency. As noted by Bunse et al. \cite{3}, the efficiency of a modern sorting algorithm is no longer defined solely by operation counts, but by memory locality and energy consumption, demanding a rigorous re-examination of classical approaches through the lens of modern micro architectures. \par
The performance of sorting algorithms is bifurcated by a theoretical limit. Comparison-based algorithms are bounded by the information theoretic lower bound. The decision tree model dictates that any deterministic comparison sort must perform $\Omega(nlogn)$ comparisons to distinguish between $n!$ permutations \cite{4}. To breach this limit, high-performance non-comparison algorithms exploit the binary representation of keys to achieve linear time $O(n.k)$, where $k$ is number of digits or bits in a key. These algorithms rely on direct indexing and arithmetic operations instead of conditional branches, resulting in fewer if-else decisions, improved CPU pipeline utilization, and superior performance on modern processors, particularly for random inputs \cite{11}. On the other hand, Comparison-based algorithms (e.g., std::sort) offer universal applicability by treating data as opaque objects as distribution-based algorithms suffer from type specialization restricting their utility to scalar data types. \par 

The current algorithmic landscape exhibits a pronounced trade-off between performance and stability. Practitioners are typically forced to choose between extreme throughput and order preservation (stability), but rarely both. Many high-throughput sorting algorithms that sacrifice stability often dominate the benchmarks, but they are insufficient for some industrial applications. For instance, in Relational Database Management Systems, stability is a correctness requirement, not a feature \cite{1}. State-of-the-art implementations such as Pdqsort (Pattern-Defeating Quicksort) \cite{11}\cite{12} and Skasort \cite{15}\cite{16} aggressively exploit instruction-level parallelism and memory bandwidth to maximize throughput. In contrast, stable sorting algorithms incur a substantial performance disadvantage due to extra memory operations. For example, in the C++ Standard Library, $std::stable\_sort$ \cite{18} is consistently observed to be significantly slower than $std::sort$ \cite{17}. Similarly, within the Boost.Sort library \cite{5}, the unstable Spreadsort outperforms the stable Spinsort across a wide range of input distributions. This persistent stability tax compels system designers to accept significant performance penalties whenever stability is a correctness requirement. \par 

Historically, Least Significant Digit (LSD) Radixsort \cite{4}\cite{19} has been the standard exempt from the stability tax, offering stable linear performance. However, we argue that this advantage is fragile in the face of modern data types. Traditional Radixsort faces a ``Bit-Width Wall" on 64-bit and larger key sizes. A byte-wise Radixsort requires only four passes for 32-bit integer keys; however, with the widespread adoption of 64-bit keys such as memory pointers and Universally Unique Identifier (UUIDs), the same approach requires eight full passes when using a conventional 8-bit radix. Although increasing the radix width can theoretically reduce the number of passes, it exponentially expands the scatter-gather range, causing TLB misses and cache thrashing that negate the benefits \cite{7}. As a result, sorting $N$ 64-bit keys under standard radix parameters effectively doubles the memory traffic incurring approximately $2N$ reads and writes per pass, and significantly amplifies cache pressure. Consequently, the throughput advantage traditionally associated with Radixsort over comparison-based algorithms diminishes as key width increases \cite{6}. \par

Modern large-scale systems have largely moved beyond 32-bit identifiers in favor of wider, collision-resistant keys. Distributed systems and NoSQL databases such as Apache Cassandra \cite{20} and MongoDB \cite{21} commonly use 128-bit UUIDs as primary keys to guarantee global uniqueness across shards without coordination \cite{8}. Furthermore, the rise of vector search and high-dimensional analytics often requires sorting 64-bit or 128-bit hash signatures (e.g., MurmurHash, CityHash) to group semantically similar entities. Consequently, sorting algorithms that were optimized for the 32-bit integer era are facing a ``width explosion", where the cost of processing a single key has doubled or quadrupled, exposing the scalability limits of traditional distribution-based sorting approaches. \par

In this work, we propose zSort, a sorting algorithm designed to reduce the stability tax. zSort introduces a novel z-score based distribution strategy that bridges the gap between the raw speed of unstable algorithms and the correctness guarantees of stable ones. Unlike traditional distribution based stable sorting techniques, which degrades linearly with key size, zSort is explicitly architected so that the complexity (number of passes) remains invariant of the key width. 

\section{Comparative Study of Existing Algorithms}
 
Comparison-based sorting algorithms are commonly implemented using merge sort based techniques such as Timsort \cite{22} and the C++ Standard Library's $std::stable\_sort$. These algorithms rely entirely on a binary comparison function $cmp(A,B)$ to determine the order between elements. Timsort incorporates several practical optimizations that allow it to exploit existing order in the input. As a result, it can achieve performance close to $O(n)$ when the data is partially sorted. However, despite these optimizations, it cannot escape the fundamental $\Omega(n \log n)$ lower bound that applies to comparison-based sorting. Another limitation arises from the merge phase of these algorithms, which requires frequent comparisons. Each comparison typically results in a conditional branch instruction. On modern superscalar processors, these branches can lead to mispredictions, causing pipeline stalls and reducing execution efficiency \cite{10}. \par 

To reduce the performance cost caused by conditional branches, recent research in algorithm engineering has explored hybrid unstable sorting algorithms that aim to increase instruction-level parallelism. Examples include Pdqsort and Block Quicksort \cite{11}. Instead of processing elements one by one, these algorithms handle data in blocks. Elements from a block are loaded into registers and compared with a pivot which helps to avoid conditional jumps. This approach allows the processor to make better use of speculative execution and keeps the CPU pipeline busy. However, this method makes the algorithm unstable. During block-based swapping, the relative order of equal elements may change. \par 

Skasort (an in-place MSD Radixsort) is a distribution based sorting technique where the data is recursively divided based on the most significant digits. This recursive partitioning gradually reduces the size of the working data, allowing more of it to fit into higher levels of the CPU cache, as discussed in \cite{13}. Because of this, Skasort benefits from better cache usage. In contrast, LSD Radixsort scans the entire array multiple times, which leads to poor temporal locality and more cache misses. By working on smaller partitions, Skasort reduces cache misses and decreases access to main memory. However, Skasort use in-place permutations that may not preserve the original order of equal keys.
 
The Boost Sort Library \cite{5} represents the collection of high performance sorting algorithms available for C++. $boost::sort::spinsort$ is a stable, adaptive sorting algorithm designed to exploit pre-existing order in the input. It combines run detection with merging and insertion-based techniques, achieving near-linear performance on nearly sorted data \cite{23}. On random input, its performance degrades to $O(nlogn)$, consistent with comparison-based stable sorts. Like $std::stable\_sort$, its worst-case complexity is bounded by the $\Omega(nlogn)$ lower bound. \par 
$boost::sort::flat\_stable\_sort$ addresses the memory overhead associated with stable sorting. Unlike $std::stable\_sort$, which typically requires $O(n)$ auxiliary memory, $flat\_stable\_sort$ preserves stability using only $O(\sqrt{n})$ additional space.
This reduction in memory usage comes at the cost of significantly increased element moves and comparisons relative to classical merge-based stable sorts \cite{24}. $boost::sort::spreadsort$ is Boost's hybrid distribution-based sorting algorithm. It employs heuristic, type-aware analysis of key characteristics and bucket sizes to dynamically select between Radixsort, and $std::sort$, enabling high performance across a wide range of data distributions \cite{25}.
Table \ref{comp_study} provides the comparative study of existing algorithms across different aspects.  
\begin{table}[!h]
	\caption{Comparative Study of Existing Algorithms}
	\label{comp_study}
	\centering
	\resizebox{\linewidth}{!}{
		\begin{tabular}{|c|c|c|c|c|c|c|}
			\hline
			Algorithm &
			\makecell{Worst-Case\\Time} &
			Locality &
			Aux Space &
			Stability &
			Adaptivity &
			Parallel \\ \hline
			Timsort    & $O(nlogn)$    & High     & $O(n)$     & Yes & High   & Low \\
			$std::stable\_sort$ & $O(nlogn)$  & Medium & $O(n)$ &Yes & Low & Low   \\
			$std::flat\_stable\_sort$ & $O(n\log n)$  & Medium      & $O(n)$       & Yes & Medium & Low \\
			Spinsort   & $O(n\log n)$    & High     & $O(n/2)$     & Yes & Medium & Low \\
			$std::sort$ & $O(nlogn)$ & High & $O(logn)$& No & Low & Medium \\
			Pdqsort    & $O(n\log n)$    & V. High  & $O(log n)$  & No  & High   & Medium \\
			Skasort  & $O(n)$          & High     & $O(k)$       & No  & Medium & Medium \\
			Spreadsort & $O(n\log n)$    & High     & $O(k)$  & No  & High   & Medium \\
			LSD Radix   & $O(n\cdot w/b)$ & Low      & $O(n+k)$       & Yes & None   & High \\
			
			\hline
		\end{tabular}
	}
\end{table}

\section{Proposed Work}
The proposed algorithm follows a distribution-based sorting paradigm in which the input problem set 
$S = \{s_1, s_2, \dots, s_n\}$ is divided into $k$ disjoint clusters. Each element $s_i \in S$ is deterministically 
assigned to exactly one cluster $C_j$, where $j \in \{0,1,2,\dots,k-1\}$, according to a predefined mapping 
function. This yields a collection of mutually exclusive subsets $C_0, C_1, \dots, C_{k-1}$ satisfying 
$C_p \cap C_q = \varnothing$ for $p \neq q$ and $\bigcup_{j=0}^{k-1} C_j = S$. \par 
The partitioning strategy enforces the ordering invariant $\max(C_j) < \min(C_{j+1})$ for all adjacent clusters, thereby guaranteeing that the clusters are totally ordered and independent. As a result, the partitioning phase produces a partially ordered structure, after which each cluster $C_j$ can be  treated as an independent subproblem and processed recursively using the same strategy. The distribution of elements into their respective clusters is governed by a deterministic mapping function, 
denoted as $h(s_i)$. This function takes an element $s_i \in S$ as input and maps it to an integer index $j \in \{0,1,\dots,k-1\}$, thereby assigning $s_i$ to the corresponding cluster $C_j$ based on its normalized position within the data distribution. The mapping function $h$ is inspired by the \emph{z-score transformation}, also known as standard scaling. The objective of this transformation is to normalize datasets with heterogeneous value ranges onto a common scale while preserving the relative ordering of elements. By maintaining order, the transformation ensures that the partitioning invariant established in the previous stage remains valid. \par
Each element $s_i \in S$ is first standardized using the z-score transformation, defined as
\[
z_i = \frac{s_i - \mu}{\sigma}, \quad i = 1,2,\dots,n,
\]
where $\mu$ and $\sigma$ denote the mean and standard deviation of the dataset, respectively. This transformation consists of two steps: \emph{centering} and \emph{scaling}. In the centering step, the mean $\mu$ is subtracted from each data point, shifting the distribution so that its new mean is zero. In the subsequent scaling step, the centered values are divided by the standard deviation $\sigma$, 
normalizing the dispersion of the data such that the resulting distribution has unit variance. \par 
After computing the z-scores for each element $s_i$, the transformed values $z_i$ may be either positive or negative. To shift the entire set into the nonnegative domain, we identify the minimum transformed value $z_{\min}$ and add its absolute value to every element. This translation preserves the relative ordering of elements while ensuring all transformed values are nonnegative.
Under the assumption of an approximately normal distribution, nearly $95\%$ of data points fall within the interval $[-2, 2]$ in z-score space. Without further refinement, this range would yield only a small number of distinct clusters. To increase the granularity of partitioning, a scaling factor is applied, effectively subdividing the transformed range into a larger number of buckets. After shifting and scaling, each value is floored to obtain an integer index $j$, which constitutes the final output of the mapping function $h$. 
The mapping function can be formally expressed as
\begin{equation}
	h(s_i) =
	\left( \frac{s_i - \mu}{\sigma} + z_{\min} \right)\cdot \text{scale}
	\label{h(si)}
\end{equation}
where $\mu$ and $\sigma$ denote the mean and standard deviation of the input set, $z_{\min}$ is the minimum 
z-score observed after standardization, and \textit{scale} is a user-defined parameter that controls the 
number of clusters. 

\begin{figure*}[!t]
	\centering
	\includegraphics[width=\textwidth]{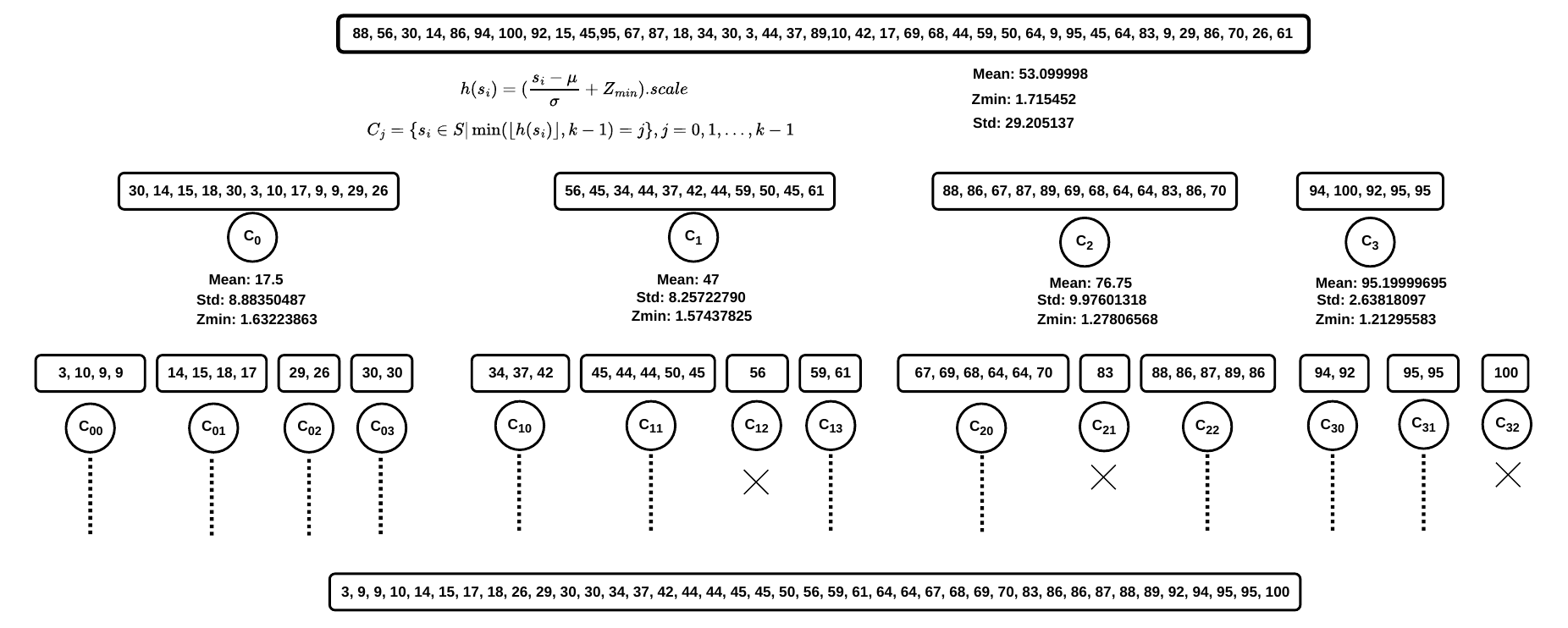}
	\caption{Sample Problem Instance}
	\label{sample}
\end{figure*}

Using this mapping, each cluster $C_j$ is defined as
\[
C_j = \left\{ s_i \in S \;\middle|\; \min\!\big(\lfloor h(s_i) \rfloor,\, k-1\big) = j \right\}, 
j = 0,1,\dots,k-1
\]
The $\min(\cdot)$ operation ensures that all mapped values are clipped to the valid cluster index range. The fundamental working of mapping and sorting process based on equation \ref{h(si)} is shown in Figure \ref{sample}. For the sample example, the number of clusters $(k)$ and $scale$ are considered to be 4 and 1 respectively. Also, the minimum problem size is considered as 1. 
The selection of $k$ is extremely important as it directly affects memory utilization and depth of the recursion tree, A heuristic approach is opted with $k = \sqrt{n}$, where $n$ denotes the size of the original problem set. Importantly, the value of $k$ is kept constant across all recursive invocations. This design choice fixes the recursion depth to two levels, since recursively partitioning a set of size $n$ into $\sqrt{n}$ clusters yields subproblems of size $\sqrt{n}$, which are resolved in the subsequent level. 
This approach is preferred over a single-level partitioning strategy with $k = n$, which is highly sensitive to skewed or non-uniform data distributions and may result in severe load imbalance across clusters. By maintaining a bounded recursion depth and uniform cluster sizes, the proposed strategy improves robustness and performance stability. \par 
The proposed algorithm is inherently stable. Stability is guaranteed by processing elements strictly in their original input order and by avoiding any operations that could invert the relative ordering of equal keys. Consequently, for any two elements $s_i$ and $s_j$ with identical keys, where $i < j$ in the input, their relative order is preserved in the final output. Furthermore, the algorithm’s performance is invariant to the input permutation. Its core computations including the calculation of the mean, standard deviation, z-scores, and subsequent distribution into clusters are order-independent. As a result, the algorithm avoids classical worst-case scenarios associated with comparison-based sorting, such as already sorted, reverse-sorted, or nearly sorted inputs, and delivers consistent performance regardless of the initial data arrangement. \par 
Due to the deterministic nature of the mapping function $h$, duplicate elements are always mapped to the same cluster $C_j$. In certain scenarios, this may result in a cluster consisting entirely of identical elements. Such a cluster cannot be further partitioned in subsequent recursive calls, as all elements would again be 
assigned to a single cluster. Recursion additionally terminates early when a cluster contains only identical elements; in this case, the corresponding subproblem is considered solved, since a set of identical elements is already trivially sorted. This mechanism enables the algorithm to handle inputs with a high degree of duplication efficiently, avoiding unnecessary recursive work and ensuring robust performance.
The procedure of the proposed method with all heuristics and hybrid approaches is presented in Algorithm \ref{alg}.
\begin{algorithm}[!h]
	\caption{zSort$(arr[\ ],\ output[\ ], size)$}
	\label{alg}
	\begin{algorithmic}[1]
		\STATE $sum \gets arr[0], min \gets sum, a \gets 1,	d \gets 1$	
		\FOR{$i = 1\ to\ size-1$}
			\STATE $curr \gets arr[i]$
			\STATE $sum+=curr$
			\STATE $min \gets curr < min ? curr : min$
		\ENDFOR
		\vspace{0.5em}
		\STATE $mean\ \gets sum/size$ 
		\IF{$(min == mean)$}
			\STATE $return\ output \gets arr$
		\ENDIF
		\vspace{0.5em}
		\STATE $var = 0$
		\FOR {$i = 0\ to\ size-1$}
		\STATE $res \gets arr[i] – mean$
		\STATE $var \gets var +  res * res$
		\ENDFOR
		\vspace{0.5em}
		\STATE $std \gets sqrt(var / size)$;
		\STATE $Zmin \gets |\frac{min-mean}{std}|$ 
	
		\STATE $clusters \gets sqrt(size)$
		\STATE $scale \gets clusters/4$
		\STATE $counts[clusters]$, $offsets[clusters] $, $clustersArr[size]$
		
        \FOR {$i = 0\ to\ size-1$}
        	\STATE $b \gets h(arr[i])$
        	\STATE $counts[b]++$
        \ENDFOR
        \vspace{0.5em}
        \STATE $countSum = 0$
        \FOR {$i = 0\ to\ clusters -1 $}
        \STATE $offfset[i] = countSum$
        \STATE $countSum \gets countSum + counts[i]$  
        \ENDFOR
        \vspace{0.5em}
		\FOR {$i = 0\ to\ size-1$}
			\STATE $b \gets h(arr[i])$
			\STATE $clustersArr[offsets[b)]++]$
		\ENDFOR
		\vspace{0.5em}
		\STATE $lbSum = 0$
		\FOR {$i = 0\ to\ clusters -1 $}
			\STATE $totalElements = counts[i]$
			\vspace{0.5em}
			\IF {$(totalElements > 0)$}
				\IF{$(totalElements > 96)$}
				\vspace{0.5em}
					\STATE $meanEst \gets (((\frac{i+0.5}{scale} - zmin) * std) + mean)$
					\vspace{0.5em}
					\STATE $zsortRec(clustersArr, output, lbSum,$\\ $totalElements, clusters, scale, meanEst$
					\vspace{0.5em}
				\ELSE 
					\STATE $InsertionSort(clustersArr, output,$\\
					\hspace{2cm}$ lbSum, totalElements)$
				\ENDIF
				\STATE $lbSum += totalElements$
					 
			\ENDIF
		\ENDFOR
			\end{algorithmic}
\end{algorithm}
In the algorithm, before distributing the elements into the clusters, a histogram is made of the clusters. This helps in maintaining one single array for all clusters separated by offsets, which improves the cache efficiency and significantly reduces the impact of the random memory access.\\
In the recursive phase $zsortRec()$, an optimization is introduced to eliminate the computational overhead with the calculation of mean. The Calculation of the exact mean requires $O(n)$ time, in order to reduce this to $O(1)$, we approximate the mean using the mid-point of the cluster's range. This method assumes that the data is uniformly distributed within the micro-range of a single cluster, allowing us to estimate the mean as the value mapping to the mid-point of the cluster. This mean is passed as an input to the recursive call, rest of the logic is exactly the same as zSort. In other words, the estimated mean is simply the mid-point of the range of the cluster that is being recursively called. This allows us to reduce one $O(n)$ loop in further calls.\\
The range of the $C_j$ is defined as $[j,j+1)$. Let $x$ and $y$ be the lower boundaries of the clusters $j$ and $j + 1$ respectively, such that:
$h(x) = j, h(y) = j + 1$
$$
\implies h(meanEst) = \frac{h(x) + h(y)}{2} = j + 0.5
\label{Estmean}
$$
We define the estimated mean $(meanEst)$ , as the inverse of the range's mid-point  using equation \ref{h(si)},
$$meanEst = ((\frac{j+0.5}{scale} - Zmin) * std) + mean$$ 
Further, few heuristic solutions are also applied to improve the performance such as switching to insertion sort when the sub problem size is less than 96 and switching to counting sort when the range of the problem is less than 65000. Also, while theoretical analysis suggests that the optimal number of clusters is $k = \sqrt{n}$, this derivation typically assumes negligible access costs and cluster management. To refine this for practical implementation, we introduce a scaling factor $\alpha$ to the cluster count, such that the number of clusters is given by $k = \alpha \times \sqrt{n}$. The sensitivity analysis is done to observe the impact of $\alpha$ on the total execution time (Table \ref{sens_analysis}). The value of $\alpha$ is varied from 0.4 to 1.2 to cover the range surrounding the theoretical optimum $(\alpha = 1.0)$. The experimental data reveals a convex optimization curve. While the theoretical baseline of $\alpha =1.0$ yields a performance of 240.80 ms, reducing the number of clusters improves performance significantly up to a global minimum at $\alpha =0.6\ (213.73 ms)$. This represents an approximate 11.2\% speedup over the theoretical square-root baseline. By reducing the bucket count to 60\% of $\sqrt{n}$, the average load per cluster increases slightly. However, this allows better CPU cache locality and reduces the metadata overhead of the cluster structure itself. The reduced memory footprint of the cluster array likely results in fewer TLB (Translation Lookaside Buffer) misses.

\begin{table}[h]
	\caption{Sensitivity Analysis for $\alpha$}
	\label{sens_analysis}
	\centering
	\resizebox{\linewidth}{!}{
	\begin{tabular}{|c|c|c|}
		\hline
		\textbf{Scaling Factor ($\alpha$)} & \textbf{Avg. Execution Time (ms)} & \textbf{Relative Performance} \\ \hline
		1.2                & 258.83                   & -17.40\%             \\
		1                  & 240.8                    & Baseline             \\
		0.8                & 214.37                   & 10.90\%              \\
		0.6                & 213.73                   & 11.20\%              \\
		0.4                & 224.32                   & 6.80\%              \\ \hline
	\end{tabular}}
\end{table}
\subsection{Complexity Analysis}
\subsubsection{Worst Case}
Let algorithm operates on input size $n$. The partition step takes $O(n)$ time. Algorithm divides the input into $k=\sqrt{n}$ parts in each level. Here, $k$ is computed from original input size $n$ and remain fixed throughout recursion. Subproblems of size $\leq t$ (Here, $t\leq 96$) are solved in $O(1)$.
Therefore, the classical recurrence is given by, $$T(n) = kT(\frac{n}{k}) + n$$
As branching factor $k$ does not depend on current subproblem size,  the depth behavior changes fundamentally. Also size of each partition $\neq \frac{n}{k}$ due to \textbf{skewness}. Let subproblem sizes after partition be - $n_1, n_2,\dots,n_k$ such that $\sum_{i=1}^{k}n_i = n$. 
Thus the recurrence now have following form, 
$$T(n) = \sum_{i=1}^{k}T(n_i)+cn$$, where $k=\sqrt{n}$ is fixed across recursion levels.  \\
Worst Case recursion Depth ($d$): We know that at every level, $\sum n_i = n$, so work per level = $\Theta(n)$ - irrespective of the skewness. Consider the worst case where one partition receives almost all the elements and remaining partitions contain a few elements (which can be considered as constant). Thus, at level 1, this partition $n_i$ will have $n - O(1)$ elements. At next level 2, this is again divided into $k=\sqrt{n}$ parts. Maximum size partition at this level will have $\leq \frac{n}{\sqrt{n}} \leq \sqrt{n}$ elements. Finally, at level 3, this is again divided in $\sqrt{n}$ part. Therefore, maximum size partition after level 3 $\leq \frac{\sqrt{n}}{\sqrt{n}}=1$. Thus recursion depth ($d\leq 3$). More precisely, $d \leq \lceil \frac{logn}{log\sqrt{n}}\rceil = \lceil 2 \rceil$, which is constant. 
Thus, total cost, $$T(n) = \sum_{i=0}^{d}cn \leq cn+cn+cn \leq 3cn = \Theta(n)$$
\textbf{General Proof:} Let $k= n^\beta$, where $0<\beta<1$. At level 1, largest size $\leq \frac{n}{n^\beta} = n^{1-\beta}$. Therefore, after $j^{th}$ level, when recurrence stops, largest size partition will be $n^{1-j\beta}\leq 1$. 
Taking log on both sides, 
$$\implies 1-j\beta = 0 \implies j \geq \frac{1}{\beta} \implies j = O(\frac{1}{\beta})$$
Thus for any value of $0<\beta<1$, depth ($j$) will be always constant. 
\subsection{Best Case}
The classical recurrence is given by, $$T(n) = kT(\frac{n}{k}) + n$$
where $k=\sqrt{n}$ is fixed across recursion levels.\\
Recursion Depth ($d$):\\
$$\frac{n}{k^d} = 1 \;\Rightarrow\; k^d = n \;\Rightarrow\; d = \log_k n\Rightarrow\;d = \log_{\sqrt{n}} n = 2$$
Here, work per level is $\Theta(n)$
Therefore, total cost $$T(n) = (\text{depth}) \times (\text{work per level}) = 2 \cdot n = \Theta(n)$$

\section{Results and Discussion}
\subsection{Experimental Setup}
The experiments are carried out on a system equipped with an ${11^{th}}$ Gen Intel Core $i5-1155G7$ processor (Tiger Lake microarchitecture) operating at a base frequency of $2.50 GHz$, coupled with $8 GB$ of Dual-Channel RAM. The system operated under the Windows environment. To ensure a fair and competitive, all algorithms are implemented in $C++$ and compiled using the $GCC\ 14.2.0$ compiler. The compilation utilized the C++ 20 standard $(-std=c++20)$ with aggressive optimization flags enabled ($-Ofast$, $-flto$, $-march=native$). 
Microarchitectural behavior is analyzed using Intel VTune Profiler 2025.7 \cite{26}. The default collection settings are used to obtain the top-down pipeline breakdown, including metrics such as Retiring, Front-End Bound, Back-End Bound, and Bad Speculation \cite{27}. Profiling is conducted at the application level, where each sorting algorithm is executed as a separate task.
\subsection{Microarchitecture Analysis}
To better understand the performance of zSort beyond its theoretical complexity, a detailed analysis of the CPU pipeline is performed. This analysis shows that the performance of sorting algorithms on modern superscalar processors depends on how the instruction stream interacts with key hardware components such as the Branch Target Buffer (BTB), Translation Lookaside Buffer (TLB), and the cache hierarchy.
Table \ref{microarch_analysis} presents the microarchitectural performance metrics observed for different sorting algorithms.
\begin{table*}[]
	\caption{Microarchitecture Analysis}
	\label{microarch_analysis}
	
	\resizebox{\linewidth}{!}{
		\centering
		\begin{tabular}{|c|cccc|cc|cc|cc|}
			\hline
			& \multicolumn{4}{c|}{\textbf{Comparison (Stable)}} 
			& \multicolumn{2}{c|}{\textbf{Comparison (Unstable)}} 
			& \multicolumn{2}{c|}{\textbf{Non-Comparison (Unstable)}} 
			& \multicolumn{2}{c|}{\textbf{Non-Comparison (Stable)}} \\ \cline{2-11}
			
			\multirow{-2}{*}{\textbf{Metric}} 
			& \textbf{Timsort} 
			& \textbf{std::stable\_sort} 
			& \textbf{flat\_stable\_sort}
			& \textbf{Spinsort}
			& \textbf{std::sort} 
			& \textbf{Pdqsort} 
			& \textbf{Skasort} 
			& \textbf{Spreadsort} 
			& \textbf{Radixsort} 
			& \textbf{zSort} \\ \hline
			
			\textbf{IPC} 
			& 0.93 & 0.69 & 1.03 & 0.82 
			& 0.58 & 2.15 
			& 1.36 & 1.26 
			& 1.25 & 1.44 \\ \hline
			
			\textbf{Frontend} 
			& 20.90\% & 20.50\% & 17.90\% & 21.20\% 
			& 20.70\% & 14.20\% 
			& 18.50\% & 8.70\% 
			& 6.10\% & 18.30\% \\ \hline
			
			\textbf{Bad Spec} 
			& 55.00\% & 55.20\% & 53.60\% & 55.30\% 
			& 55.90\% & 25.70\% 
			& 23.10\% & 46.30\% 
			& 39.90\% & 19.70\% \\ \hline
			
			\textbf{Retiring} 
			& 18.70\% & 18.90\% & 22.60\% & 18.90\% 
			& 17.70\% & 52.40\% 
			& 46.20\% & 32.10\% 
			& 28.60\% & 46.70\% \\ \hline
			
			\textbf{Backend} 
			& 5.50\% & 5.40\% & 5.90\% & 4.60\% 
			& 5.70\% & 7.80\% 
			& 12.20\% & 12.90\% 
			& 25.40\% & 15.30\% \\ \hline 
			
			\textbf{Memory Bound} 
			& 8.70\% & 9.00\% & 9.10\% & 8.00\% 
			& 10.10\% & 11.20\% 
			& 19.10\% & 27.60\% 
			& 35.80\% & 24.50\% \\ \hline
			
			\textbf{L1 Bound} 
			& 1.70\% & 0.20\% & 3.20\% & 3.00\% 
			& 3.60\% & 0.20\% 
			& 0.00\% & 7.40\% 
			& 0.00\% & 3.60\% \\ \hline
			
			\textbf{L2 Bound} 
			& 0.20\% & 0.30\% & 0.60\% & 0.40\% 
			& 2.10\% & 5.00\% 
			& 10.50\% & 3.10\% 
			& 7.40\% & 8.90\% \\ \hline
			
			\textbf{L3 Bound} 
			& 0.90\% & 2.20\% & 1.30\% & 1.10\% 
			& 0.90\% & 0.30\% 
			& 1.00\% & 4.70\% 
			& 4.40\% & 2.00\% \\ \hline
			
			\textbf{DRAM Bound} 
			& 5.80\% & 6.20\% & 4.00\% & 3.40\% 
			& 3.50\% & 5.70\% 
			& 6.70\% & 12.40\% 
			& 12.90\% & 6.00\% \\ \hline
			
			\textbf{TLB Miss \%} 
			& 0.57\% & 0.68\% & 0.34\% & 0.39\% 
			& 0.21\% & 0.15\% 
			& 0.14\% & 2.11\% 
			& 1.58\% & 2.94\% \\ \hline
			
		\end{tabular}
	}
\end{table*}
 
\begin{itemize}
	\item \textbf{Instruction-Level Efficiency (IPC (Instructions Per Cycle) and Retiring):}
The proposed zSort shows strong instruction-level efficiency, achieving an IPC of 1.44. This value is better than traditional comparison-based algorithms such as Timsort (0.93) and $std::sort$ (0.58), and it also outperforms other non-comparison methods like Radixsort (1.25). Among all evaluated techniques, Pdqsort achieves the highest IPC, placing zSort second. However, Pdqsort does not provide stability. One of the reasons for this efficiency is the simplicity of zSort’s inner loops. The loops are straightforward and avoid complex loop-carried data dependencies. Because of this, modern compilers can effectively apply auto-vectorization using SIMD, allowing more operations to be executed per clock cycle. As a result, zSort is able to utilize CPU execution resources efficiently and maintain a high level of instruction-level parallelism. This behavior is also reflected in the retiring metric, where zSort achieves 46.70\%, the highest among all stable algorithms. Although Pdqsort records a higher retiring value of 52.40\%, it executes $69 \times 10^9$ instructions, while zSort executes only $33.06 \times 10^9$ instructions. In comparison, Radixsort (28.60\%) and Spreadsort (32.10\%) are less efficient despite being non-comparison methods. The optimized version of Radixsort (Skasort) shows a retiring percentage similar to zSort. Overall, these results indicate that zSort reduces wasted cycles and performs a larger portion of useful computation.
	\item \textbf{Frontend and Speculation Behavior:} 
	zSort maintains a front-end bound of 18.30\%, which is comparable to both comparison-based and non-comparison-based algorithms. This indicates that the instruction fetch and decode stages are not major bottlenecks, and the algorithm is able to maintain a steady flow of instructions to the execution pipeline. zSort also shows a bad speculation rate of 19.70\%, which is better than Radixsort (39.90\%) and Spreadsort (46.30\%). Fewer branch mispredictions reduce pipeline stalls and improve overall execution efficiency. Compared to unstable comparison-based methods such as Pdqsort (52.40\%), zSort provides a more predictable execution pattern while also maintaining stability. Unlike comparison-based sorting algorithms, zSort uses a deterministic mapping function $h(s_i)$ to directly map elements to their positions instead of relying on branch-heavy comparisons. This design reduces control-flow uncertainty and helps lower the bad speculation rate to 19.70\%, which is the lowest among the evaluated algorithms.
	\item \textbf{Backend and Execution Bottlenecks:} The backend bound for zSort is measured at 15.30\% which indicates moderate stalls. This value is lower than memory-intensive methods such as Radixsort (25.40\%), but it still suggests some inefficiencies in memory access. A defining characteristic of zSort is the asymmetry between its load (read) and store (write) behaviors, which directly contributes to backend performance.
	\begin{enumerate}
		\item Sequential Reading: During input processing, zSort reads data sequentially. This allows the hardware prefetcher to accurately predict upcoming memory accesses. It minimizes the backend stalls caused by load operations.
		\item Scattered Writing : In contrast, the output phase performs bucket-based distribution, which leads to scattered writes across a large virtual address space. Consequently, DTLB store misses are higher than load misses. These frequent TLB misses (2.94\%) during store operations introduce address translation overhead and stall the backend pipeline. This backend bottleneck is therefore not due to computation or read latency, but rather write-side memory irregularity and TLB pressure. 
	\end{enumerate}
	\item \textbf{Memory Behavior and Cache Utilization:} 
	Memory behavior represents how CPU cycles are distributed across different levels of the memory hierarchy, indicating where delays occur due to data access and how efficiently memory resources are utilized during execution. zSort exhibits a memory bound of 24.50\%, which is higher than comparison-based algorithms (typically below 12\%) and Skasort (19.10\%). This is due to fact that non-comparison based stable sorting algorithms inherently rely more on memory operations and requires to allocate the auxiliary space to maintain stability. Other non-comparison stable algorithm like Radixsort is more memory bounded (35.80\%) than the zSort algorithm. This behavior is expected, as recursive partitioning behavior of zSort progressively reduces the working set which will eventually fit into the cache, leads to improved locality and reduced DRAM traffic. 
	At the L1 cache level, zSort shows a bound of 3.60\%, indicating only limited pressure on the fastest cache. This occurs because some phases of the algorithm involve irregular memory access patterns (distribution phase). At the L2 cache level, zSort records the highest bound (8.90\%) among all evaluated techniques. This suggests that many memory accesses are served from L2. In other words, the working set often exceeds the capacity of L1 but still fits within L2. This behavior is beneficial because accessing L2 is much faster than accessing main memory. At the L3 cache level, zSort shows a bound of 2.00\%, indicating that the algorithm uses deeper cache levels in a controlled manner without relying heavily on the last-level cache. This pattern is consistent with the low DRAM bound, showing that zSort successfully reduces expensive main memory accesses by making effective use of the cache hierarchy. Overall, zSort trades slightly higher L2 usage for fewer DRAM accesses, which improves overall performance. This cache behavior is largely influenced by the difference between the read and write access patterns in zSort. Sequential reads exhibit high spatial locality, allowing efficient cache utilization and minimal stall overhead. In contrast, scattered writes reduce locality, increases L2 pressure and leads to the memory stalls. However, optimizing data locality could further reduce L2 pressure and enhance cache efficiency.	
\end{itemize}
\subsection{Experimental Results}
To analyze the robustness of the sorting algorithms, the synthetic datasets representing various real-world data characteristics are considered. The data consisted of 64-bit signed integers generated using the $std::mt19937\_64\ Mersenne\ Twister\ engine$, with input sizes $n$ varying from $10^5$ to $10^7$ elements. A comprehensive evaluation is done under varying entropy and structural conditions using 5 distinct test cases consisting of Uniform distribution, Normal distribution, Skewed distribution, Nearly Sorted data and High duplicate data. Normal distribution is  modeled via a Gaussian function $N(\mu,\sigma)$ and Skewed data is generated following a power-law distribution to simulate heavy-tailed workloads. For nearly sorted data, sequences created by subjecting ordered arrays to a fixed percentage of random swaps and High-Duplicate collections drawn from a restricted value range.
Timing is measured using $std::chrono::steady\_clock$. To ensure statistical stability and minimize noise, each experiment is repeated 100 times, and the average execution time is considered. To simulate a realistic ``cold" start for every sort operation, the cache-flushing mechanism is implemented. In this, before every individual sort execution, a 128 $MB$ buffer (exceeding the CPU's L3 cache size) is explicitly traversed and modified to invalidate existing cache lines. It ensures the input data is fetched from main memory rather than residual cache.
\subsubsection{\textbf{Uniform Distribution}}
\begin{figure}[!h]
	\centering
	\includegraphics[scale= 0.4]{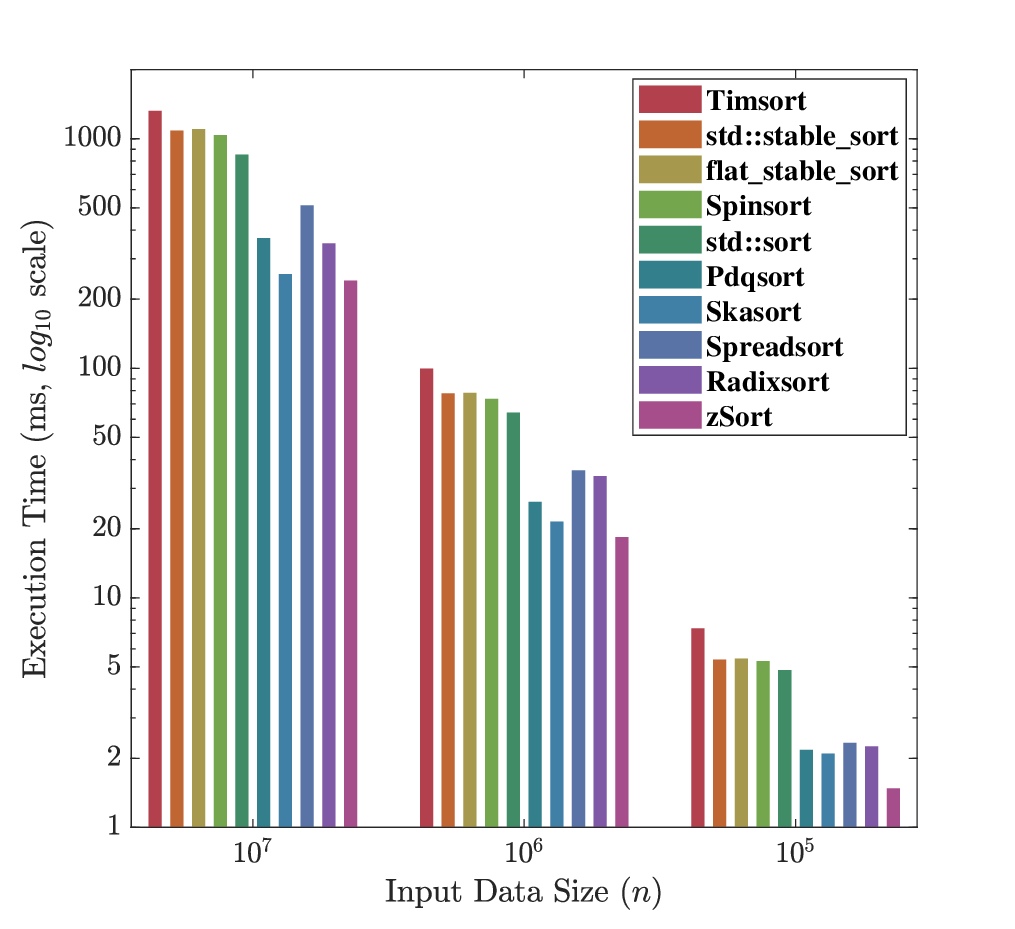}
	\caption{Execution Time vs Input Data Size under Uniform Distribution}
	\label{uniform}
\end{figure}

The performance of various sorting techniques for a uniform data distribution across input sizes of 
$10^5, 10^6$ and $10^7$ is illustrated in Figure \ref{uniform}. The uniform distribution corresponds to a high-entropy workload and serves as a baseline for evaluating general-case performance. As shown, zSort consistently achieves the lowest execution time across all input sizes, recording 1.48 ms, 18.41 ms, and 240.68 ms for input size $10^5, 10^6$ and $10^7$ respectively. A key observation is the elimination of the stability tax. At $n=10^7$, zSort significantly outperforms $std::stable\_sort$ (1086.02 ms) and $flat\_stable\_sort$ (1101.28 ms), achieving approximately a 4.5$x$ improvement. It also outperforms Spinsort (1036.71 ms), showing that its distribution-based design avoids the $O(nlogn)$ limitation and reduces the branch misprediction overhead. Furthermore, zSort consistently performs better than several high-performance unstable algorithms. At $n = 10^7$, zSort achieves about 6-35\% improvement compared to Skasort (256.88 ms) and Pdqsort (369.14 ms). This result challenges the traditional trade-off between stability and peak performance. It shows that preservation of the order of equal elements is possible without losing efficiency. When compared with radix-based approaches such as Spreadsort (512.72 ms) and Radixsort (350.27 ms), zSort achieves performance improvements of about 53\% and 31\%, respectively. Similar trends are also observed for smaller input sizes, demonstrating the robustness and scalability of zSort for large-scale 64-bit key sorting.
\subsubsection{\textbf{Normal Distribution}}  
\begin{figure}[h]
	\centering
	\includegraphics[scale= 0.4]{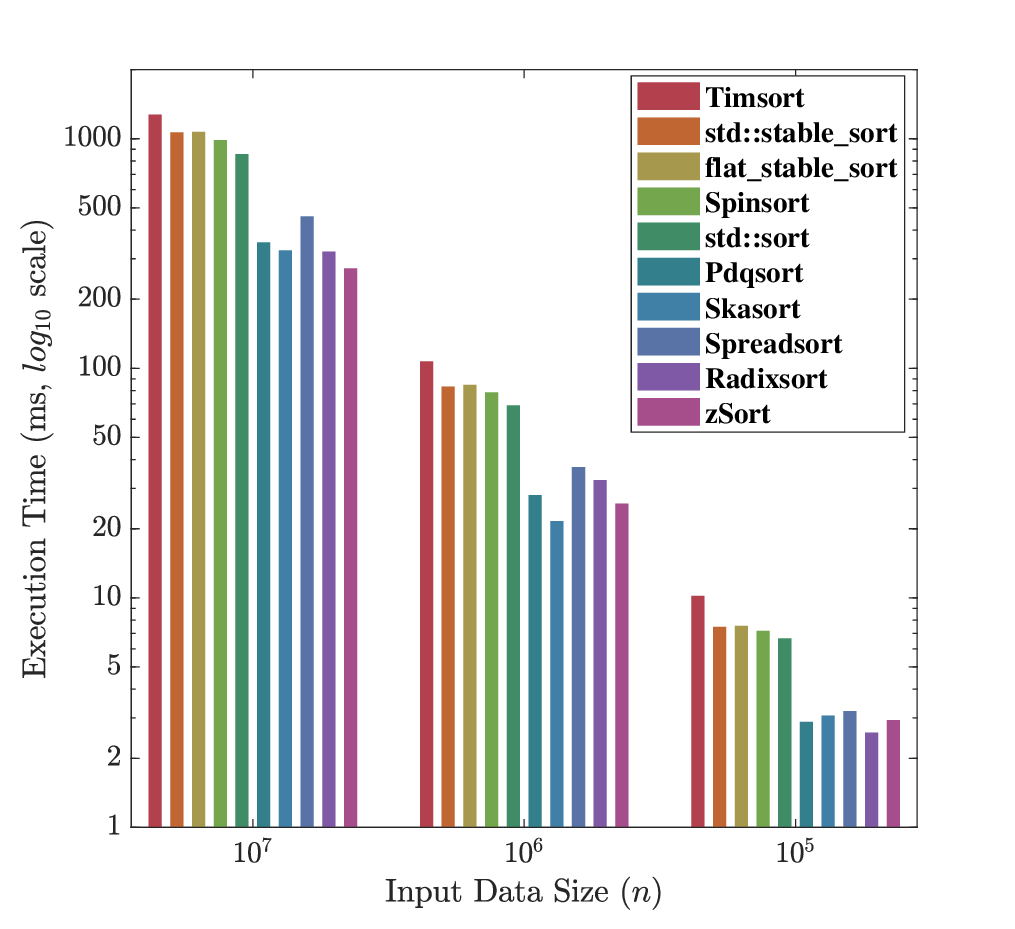}
	\caption{Execution Time vs Input Data Size under Normal Distribution}
	\label{normal}
\end{figure}

The performance of different sorting techniques under a normal (Gaussian) input distribution is shown in Figure \ref{normal}. A Gaussian distribution has a high concentration of keys around the mean, which leads to localized clustering. Such clustering can create load-balancing issues for simple bucket-based methods and can also affect pivot selection in comparison-based algorithms. Despite these challenges, zSort achieves high performance (272.37 ms) for an input size of $10^7$. In comparison, standard stable sorting algorithms such as $std::stable\_sort$ (1065.18 ms) and Spinsort (986.83 ms) are almost 4$x$ slower. Among unstable comparison based algorithms, Pdqsort performs relatively well (359.94 ms) due to its improved pivot sampling strategy. However, non-comparison-based approaches, including Skasort (326.16 ms), Spreadsort (359.37 ms), and Radixsort (322.46 ms), exhibit noticeable performance degradation compared to their uniform distribution behavior. These results indicate that zSort's recursive mapping strategy is highly resilient to localized data clustering.
\subsubsection{\textbf{Skewed Distribution}}
\begin{figure}[h]
	\centering
	\includegraphics[scale= 0.4]{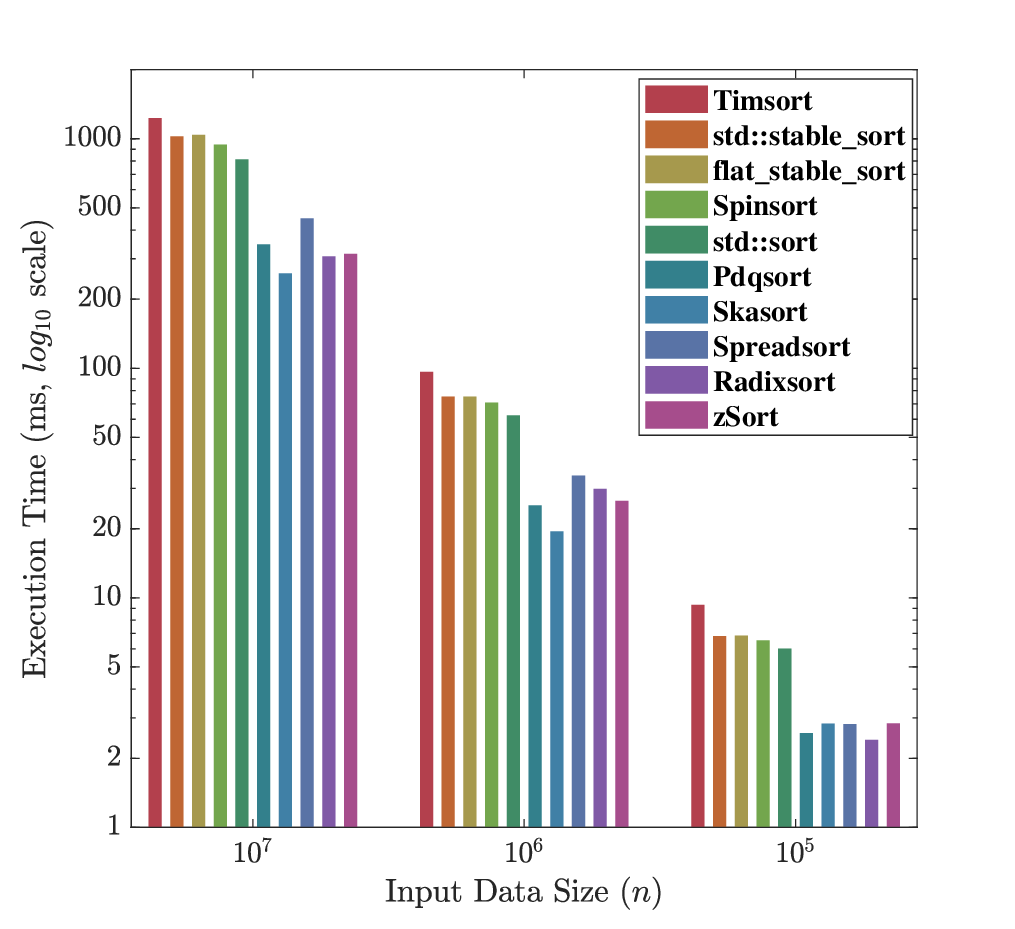}
	\caption{Execution Time vs Input Data Size under Skewed Distribution}
	\label{skewed}
\end{figure}

Figure \ref{skewed} shows the performance of various sorting techniques under a skewed input distribution.  Skewed dataset modeled using power-law distributions, represent heavy-tailed workloads where a small subset of keys dominates the frequency. Such distributions can significantly stress partition balance and increase recursion depth in both comparison-based and distribution-based algorithms. Under this distribution, zSort records an execution time of 315.13 ms indicating a slight performance degradation compared to uniform and normal distributions due to the overhead associated with handling highly skewed key distributions.
Despite this, zSort continues to outperform all stable sorting algorithms by a significant margin. Compared to Spinsort (942.16 ms) and $std::stable\_sort$ (1024.52 ms), zSort achieves approximately 3$x$ and 3.25$x$ speedups, respectively. Furthermore, zSort remains highly competitive with non-comparison-based methods, operating within approximately 2.6\% of Radixsort (307.64 ms).
\subsubsection{\textbf{Nearly Sorted Data}}
\begin{figure}[!h]
	\centering
	\includegraphics[scale= 0.4]{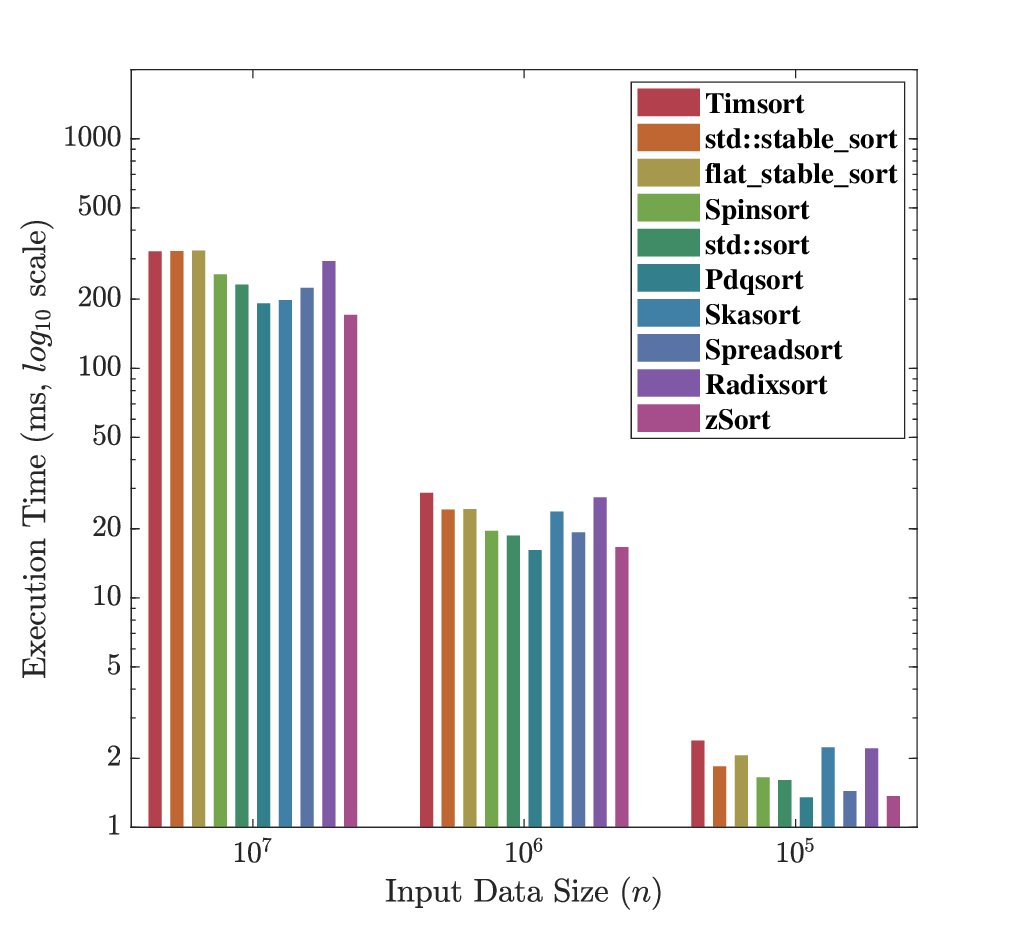}
	\caption{Execution Time vs Input Data Size under Nearly Sorted Data}
	\label{nearly_sorted}
\end{figure}

The performance of different sorting algorithms under a nearly sorted input distribution is shown in Figure \ref{nearly_sorted}. Algorithms such as Timsort, Spinsort, and Pdqsort are designed to take advantage of this type of input. However, their practical performance is still affected by microarchitectural factors. For instance, during the merge phase the frequent conditional comparisons lead to branch mispredictions.
In contrast, zSort provides better execution time compared to all other techniques across different input data sizes. This is due to the fact that zSort relies on direct memory mapping, which converts the near-sorted structure into predictable memory access patterns. Among unstable algorithms, Pdqsort performs competitively (191.56 ms) due to insertion-sort optimizations for nearly sorted data. These results show that zSort can maintain high performance even in cases where adaptive algorithms are expected to perform well.
\subsubsection{\textbf{High Duplicate Data}}
\begin{figure}[h]
\centering
\includegraphics[scale= 0.4]{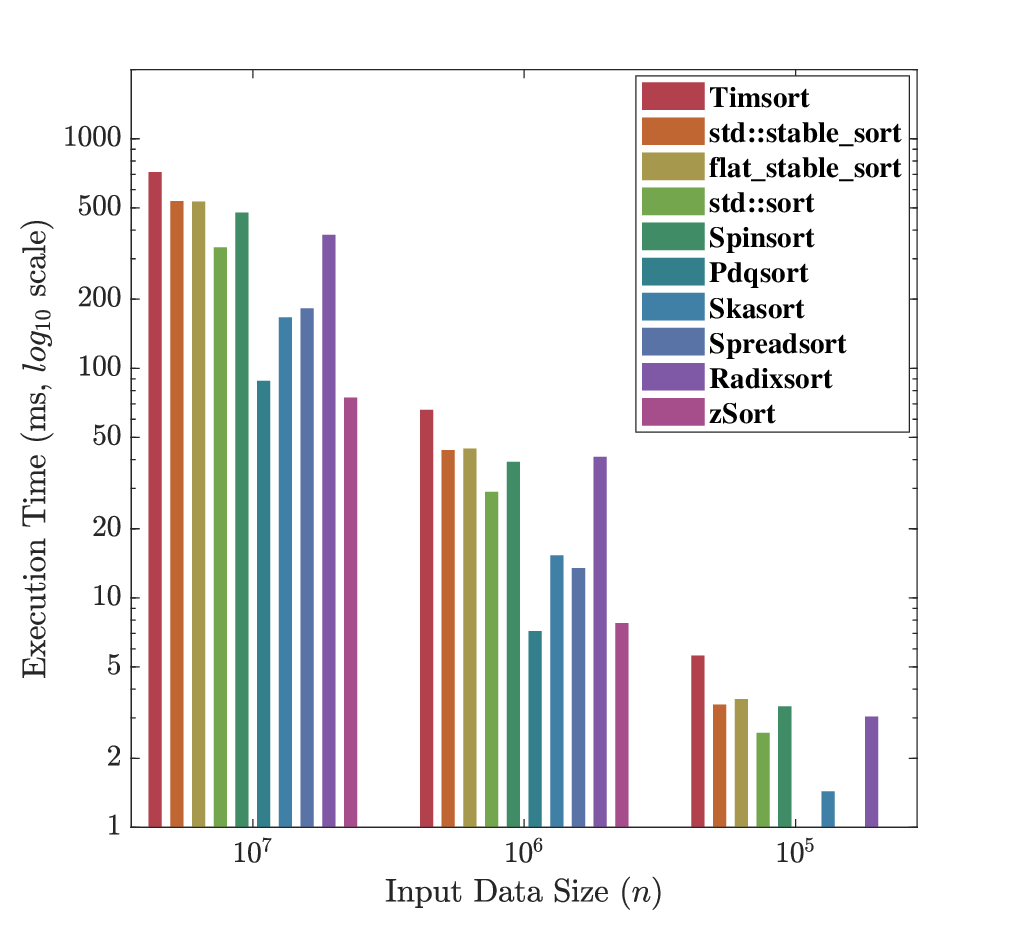}
\caption{Execution Time vs Input Data Size under Duplicate Data}
\label{duplicates}
\end{figure}

The performance of different sorting algorithms under a high-duplicate input distribution is shown in Figure \ref{duplicates}. Such datasets are challenging for both comparison-based and distribution-based algorithms because many elements have identical values, which can lead to redundant operations. Traditional algorithms like Timsort and $std::stable\_sort$ perform many repeated comparisons, resulting in slower execution times (716.05 ms and 534.99 ms, respectively). Similarly, Radixsort does not benefit from duplicate values because it processes all digit positions in the same way. In contrast, Pdqsort can handle duplicates more efficiently by using duplicate detection, achieving a time of 88.18 ms. zSort performs  better across all input sizes as compared to all other techniques. It achieves run time of 74.54 ms, 7.77 ms and 0.50 ms at input size $10^7$,$10^6$ and $10^5$ respectively. zSort improves performance by grouping identical keys during the distribution phase, which avoids redundant scattered writes. This also improves memory locality which leads to better execution efficiency. The results show that zSort handles duplicate-heavy datasets efficiently while maintaining higher throughput than both comparison-based and traditional distribution-based sorting algorithms.
\subsubsection{\textbf{Relative Performance of Sorting Algorithms Across Input Distributions}}
\begin{figure}[h]
	\centering
	\includegraphics[scale= 0.5]{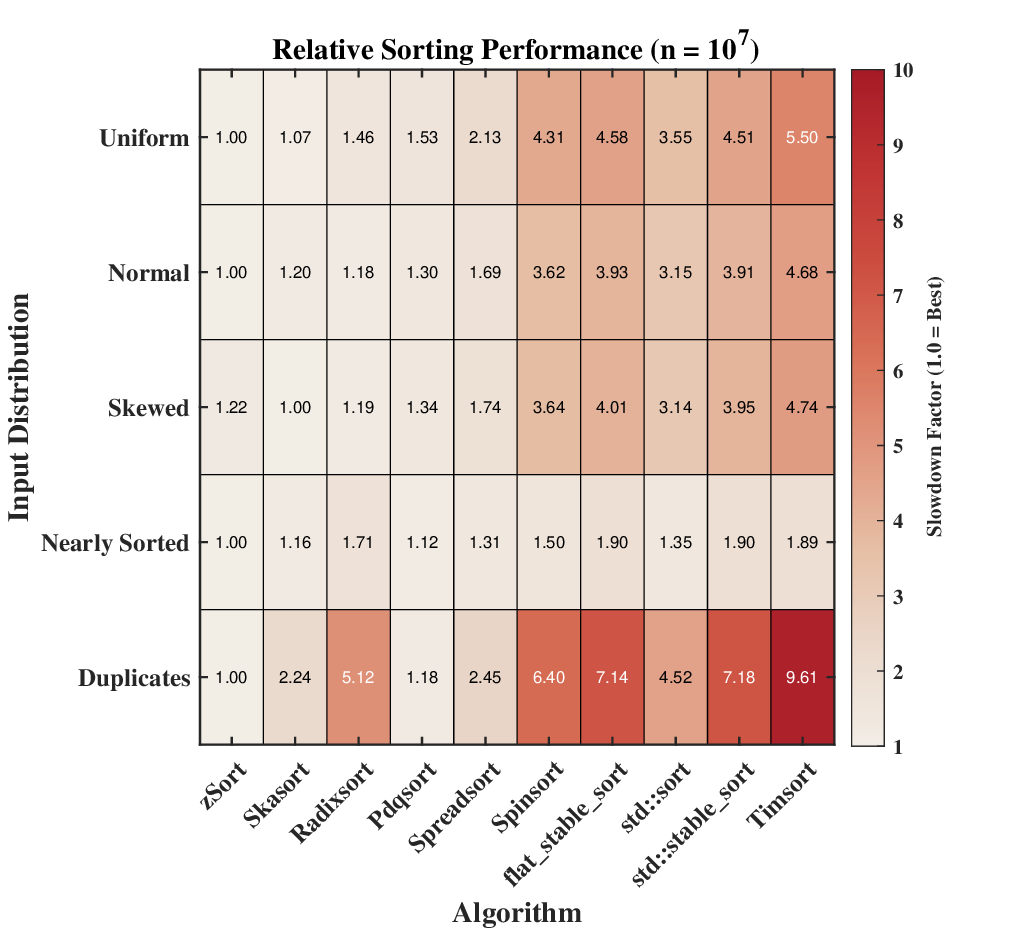}
	\caption{Relative Performance of Sorting Algorithms Across Input Distributions}
	\label{heatmap}
\end{figure}

The performance of sorting algorithms for different input distributions highlights the specialized nature of the respective existing technique. Unstable algorithms often perform very well in particular scenarios. For example, Pdqsort achieves high instruction-level efficiency on structured data, while Skasort benefits from strong cache locality in skewed data. These observations are summarized in Figure \ref{heatmap}. The results show that zSort not only leads among stable algorithms but also frequently sets the overall performance baseline. It achieves significant speedups of 3$x$ to 4.5$x$ over standard stable implementations such as $std::stable\_sort$, Spinsort, and Timsort, effectively eliminating the conventional stability overhead. 

\section{Conclusion}
In this paper the performance gap between unstable and stable sorting algorithms is addressed. While approaches such as LSD radix sort partially close this gap, their performance remains sensitive to key width due to multiple passes over the data. The paper proposes a novel sorting algorithm, zSort, that bridges this performance gap while maintaining pass complexity that is invariant to key width. zSort employs a z-score based distribution hybrid recursive structure that is well aligned with modern processor, as demonstrated through the micro-architectural analysis. Experimental evaluation shows that zSort achieves better performance across a wide range of datasets, key distributions, and input sizes as compared to the other existing algorithms. The results suggest that zSort provides a practical and efficient alternative for stable sorting in modern data processing environments. Future work includes extending the proposed algorithm to fully exploit modern hardware platforms, including multi-core CPUs and GPUs, as well as adapting it to external-memory and distributed architectures. 
\bibliographystyle{unsrt}
\bibliography{sample}
\begin{IEEEbiography}[{\includegraphics[width=1in,height=1.25in,clip,keepaspectratio]{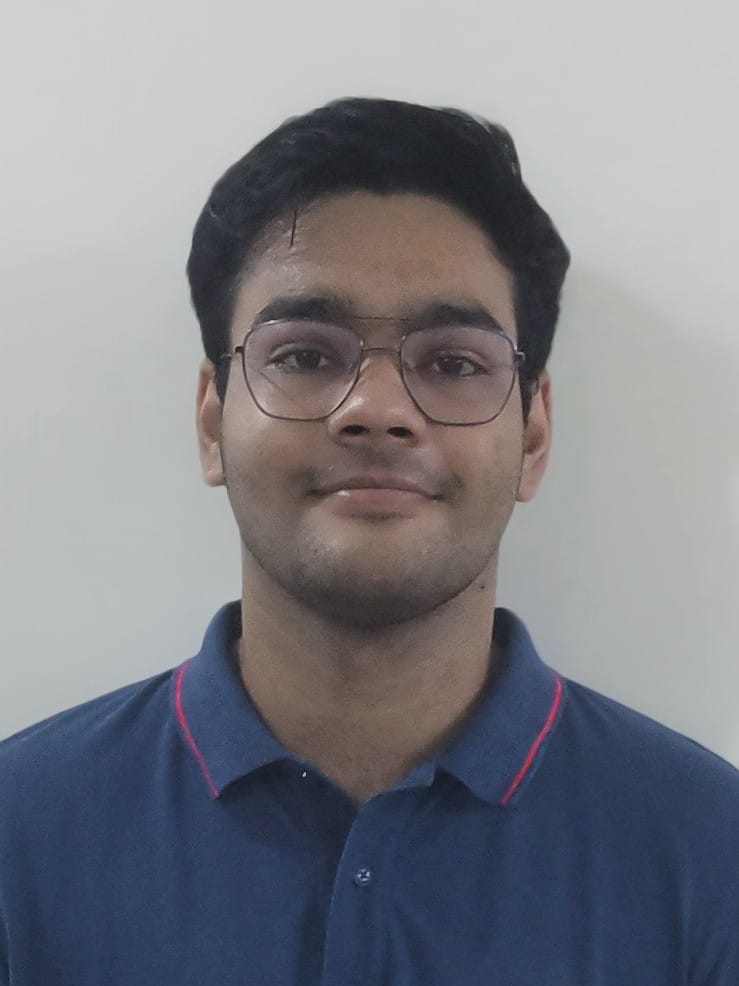}}]{Hriday Jain} is pursuing a Bachelor of Technology (B.Tech) in Computer Science and Engineering at the School of Technology, Pandit Deendayal Energy University, Gandhinagar, Gujarat, India. His research interests include algorithms, data structures, and high-performance computing. He is currently working on developing a novel sorting algorithm comparable to industry-standard methods and exploring innovative data structures using unconventional approaches.
\end{IEEEbiography}
\begin{IEEEbiography}[{\includegraphics[width=1in,height=1.25in,clip,keepaspectratio]{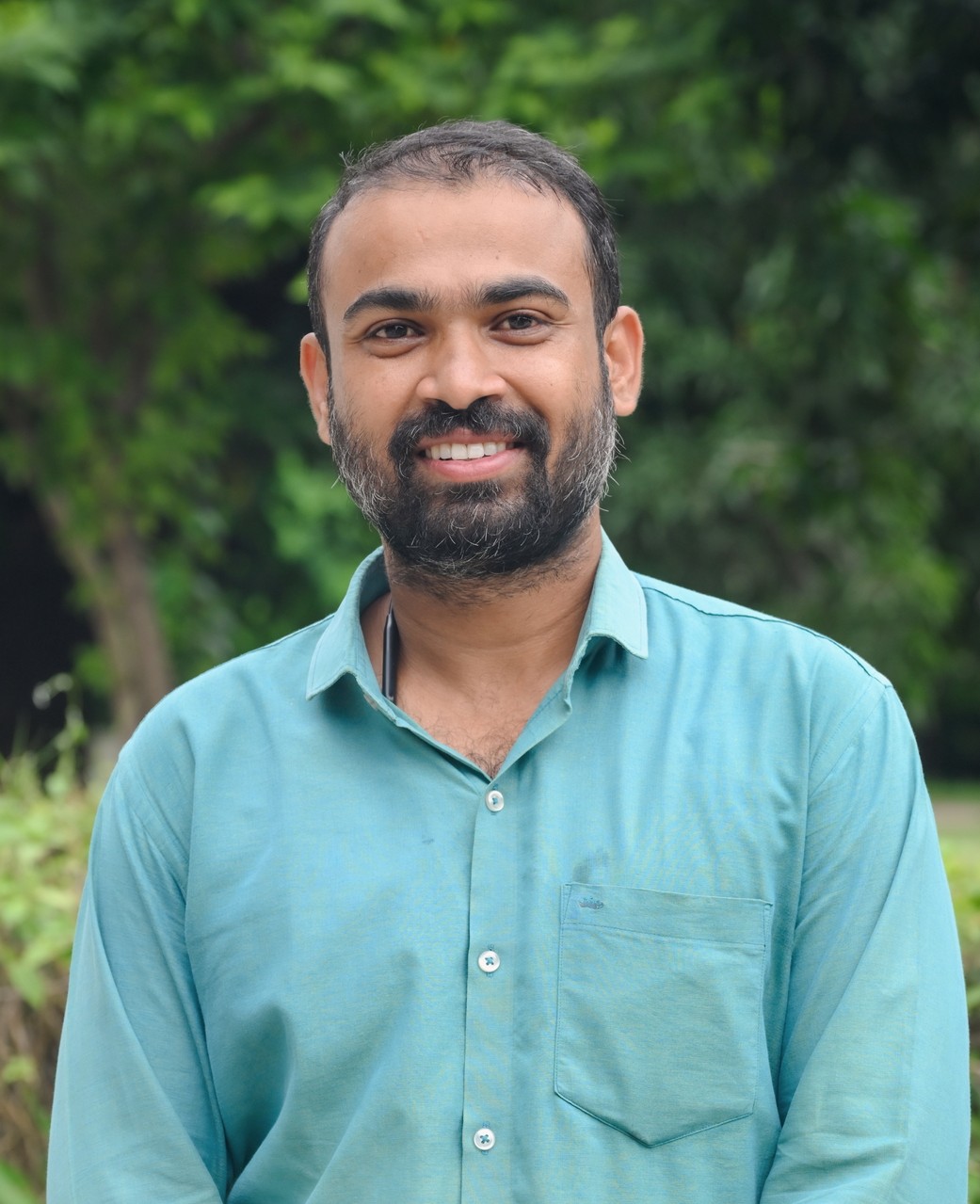}}]{Ketan Sabale} received his M.Tech. (2016) and Ph.D. (2022) degrees in Computer Science and Engineering from the National Institute of Technology, Goa. He is currently an Assistant Professor with the Department of Computer Science and Engineering at Pandit Deendayal Energy University, Gandhinagar, Gujarat, India. His research interests include Wireless Sensor Networks, Mobile ad hoc Networks, Data Structures and Algorithms.
\end{IEEEbiography}
\begin{IEEEbiography}[{\includegraphics[width=1in,height=1.25in,clip,keepaspectratio]{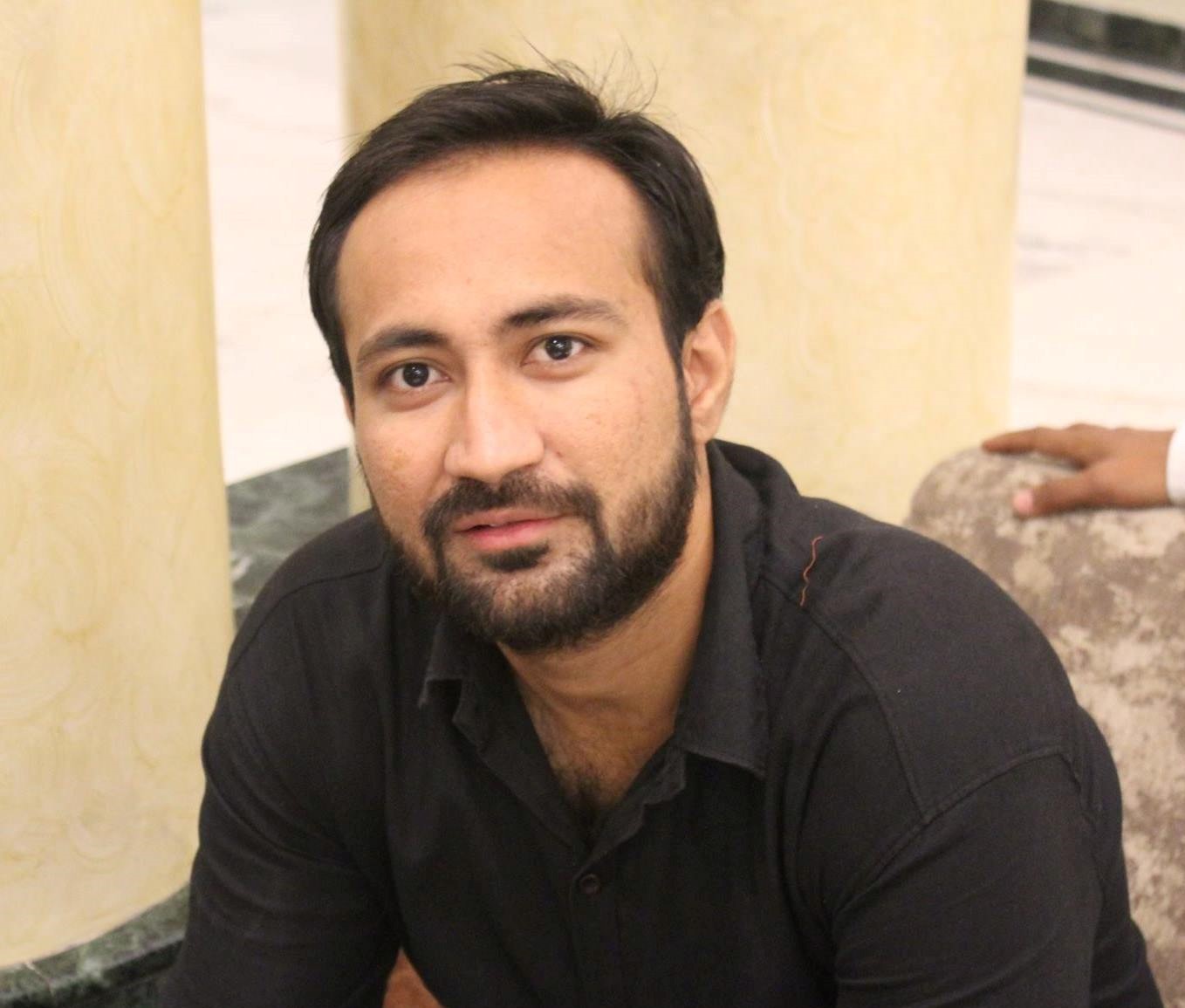}}]{Aditya Shastri} is an Assistant Professor in the Department of Computer Science and Engineering at Pandit Deendayal Energy University (PDEU), Gandhinagar, Gujarat. He received his Ph.D. in Computer Science and Engineering from the Indian Institute of Technology Indore. Broadly, he works in the field of Computational and Data Sciences. This encompasses the use of Numerical Techniques and Computer Science to solve Science and Engineering Problems. Currently, he is working on clustering and sampling techniques (Machine Learning) for Biology Applications.
\end{IEEEbiography}
\begin{IEEEbiography}[{\includegraphics[width=1in,height=1.25in,clip,keepaspectratio]{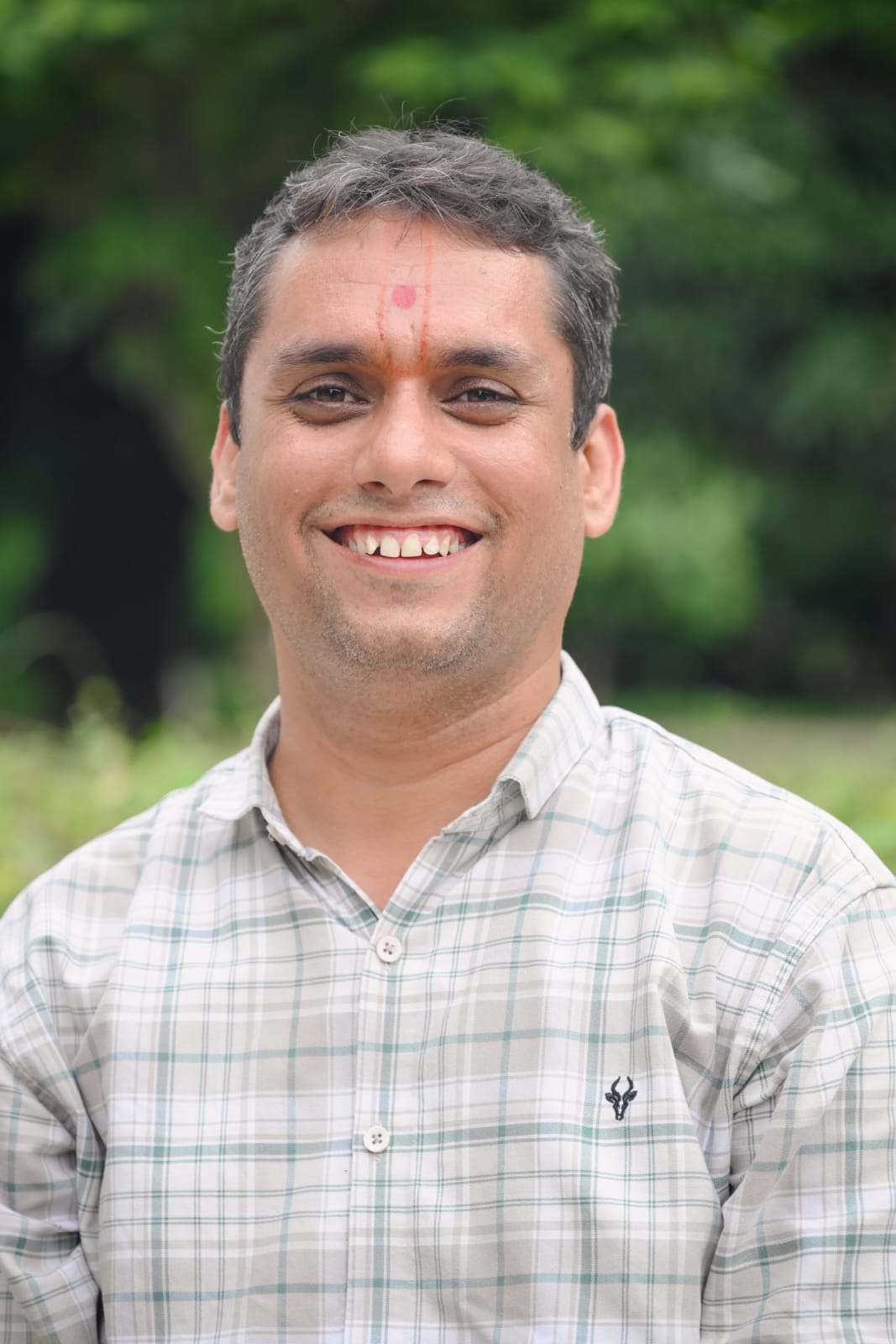}}]{Dr. Hiren Kumar Thakkar (Senior Member, IEEE) } is an Assistant Professor in the Department of Computer Science and Engineering at Pandit Deendayal Energy University, Gandhinagar, India. He received his Ph.D. in Computer Science and Information Engineering from Chang Gung University, Taiwan, in 2018, and his M.Tech. in Computer Science and Engineering from IIIT Bhubaneswar in 2012. His research focuses on biomedical signal processing, machine learning, digital health technologies, computer vision, and natural language processing. He has published extensively in reputed journals including IEEE Sensors Journal, IEEE JSAC, and IEEE Transactions on Parallel and Distributed Systems, and has contributed to several international conferences and edited books. His work emphasizes AI-driven healthcare analytics, non-invasive cardiac monitoring using ECG and seismocardiogram signals, and intelligent data-driven systems.
\end{IEEEbiography}
\begin{IEEEbiography}[{\includegraphics[width=1in,height=1.25in,clip,keepaspectratio]{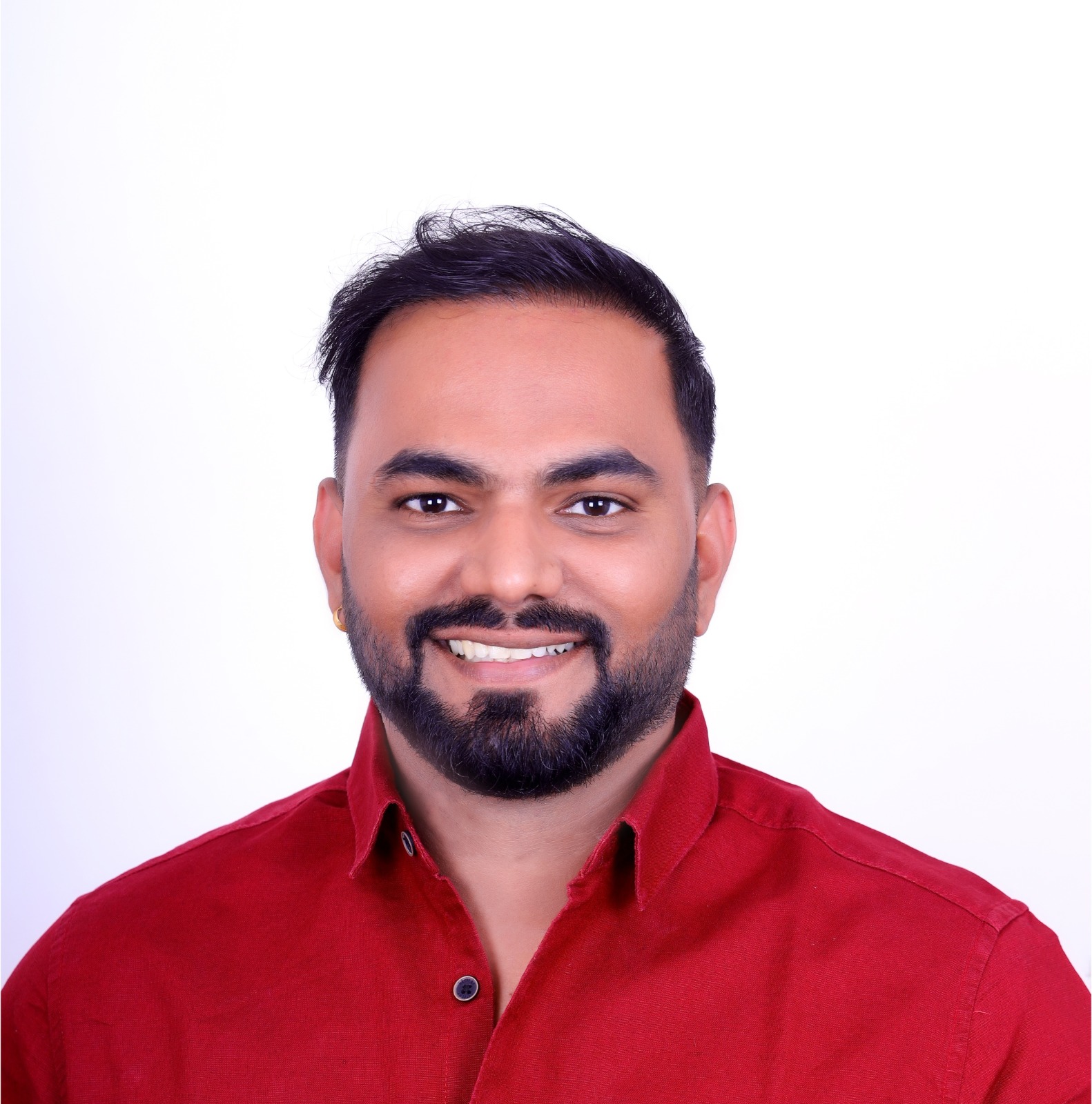}}]{Ashutosh Londhe} a Research Fellow in High-Performance Computing at the School of Mechanical and Aerospace Engineering, Queen's University Belfast. He is currently working on the GPU-eCSE02-49 project, “Very High-Order Solver Frameworks for Compressible Turbulent Mixing”, where he is enabling the FLAMENCO application to run efficiently on large-scale heterogeneous computing systems using the OPS Domain-Specific Language (DSL) framework. Previously, he was a Research Fellow at the Department of Computer Science, University of Warwick. Prior to academia, he worked at the Centre for Development of Advanced Computing (CDAC), Pune, India; Hewlett Packard Enterprise (HPE), Bangalore, India; and Advanced Micro Devices (AMD), Bangalore, India. His work included parallelisation and porting of seismic applications for large-scale heterogeneous systems, as well as application behaviour and performance analysis on Fabric Attached Memory (FAM) and Slingshot interconnect technologies, among others.
\end{IEEEbiography}
\vfill

\end{document}